# Intercomparison of the northern hemisphere winter mid-latitude atmospheric variability of the IPCC models

Valerio Lucarini[1], Sandro Calmanti[2], Alessandro dell'Aquila[2], Paolo M. Ruti[2], and Antonio Speranza[1]

[1] *Physics and Applied Statistics of Earth Fluids – PASEF - Dipartimento di Matematica ed Informatica, Università di Camerino, Camerino (MC), Italy*

[2] *Progetto Speciale Clima Globale, Ente Nazionale per le Nuove Tecnologie, l'Energia e l'Ambiente, Roma, Italy*

Corresponding author:

Dr. Valerio Lucarini

Dipartimento di Matematica ed Informatica, Università di Camerino, 62032, Camerino (MC), Italy

E-mail: valerio.lucarini@unicam.it

Tel: +39-347-6141563

Fax: +39-0737-632525

## Abstract

We compare, for the overlapping time frame 1962-2000, the estimate of the northern hemisphere mid-latitude winter atmospheric variability within the available XX century simulations of 17 global climate models included in the Intergovernmental Panel on Climate Change – 4$^{th}$ Assessment Report with the NCEP-NCAR and ECMWF reanalyses. We compute the Hayashi spectra of the 500hPa geopotential height fields and introduce an *ad hoc* integral measure of the variability observed in the Northern Hemisphere on different spectral sub-domains. The total wave variability is taken as a global scalar metrics describing the overall performance of each model, while the total variability pertaining to the eastward propagating baroclinic waves and to the planetary waves are taken as scalar metrics describing the performance of each model in describing the corresponding specific physical process. Only two very high-resolution global climate models have a rather good agreement with reanalyses. Large biases, in most cases larger than 20%, are found in all the considered metrics between the wave climatologies of most IPCC models and the reanalyses. The span of the climatologies of the various models is, in all cases, around 50% of the climatology of the reanalyses. In particular, the travelling baroclinic waves are typically overestimated by the climate models, while the planetary waves are usually underestimated. This closely resembles the results of many diagnostic studies performed in the past on global weather forecasting models. When comparing the results of various versions of similar models, it appears clear that in some cases the vertical resolution of the atmosphere and, somewhat unexpectedly, of the adopted ocean model seem to be critical in determining the agreement of the climate models with the reanalyses. The models ensemble obtained by arithmetic averaging of the results of all models is biased with respect to the reanalyses but is comparable to the best 5 models. This study suggests serious caveats with respect to the ability of most of the presently available climate models in representing the statistical properties of the global scale atmospheric dynamics of the present climate and, *a fortiori*, in the perspective of modelling climate change.

# 1. Introduction

The climate system is multi-component as well as highly non-linear: the task of planning practical strategies for improving numerical climate models is a tantalizing one. The Project for Climate Model Diagnostics and Intercomparison (PCMDI http://www-pcmdi.llnl.gov) has supported the gathering into a single location of climate model outputs contributing to the activities initiated by the Intergovernmental Panel on Climate Change (IPCC). The PCMDI thus provides a unique opportunity for evaluating the state-of-the-art capabilities in simulating the behaviour of climate system.

In particular, the improvement of diagnostic tools employed in modelling studies and the provision of simple scalar metrics of model performances is solicited by the PCMDI. Examples of scalar metrics are shown in the IPCC-Third Assessment Report (hereafter TAR), chapter 8, in the form of correlations between observed and simulated quantities, relative amplitude of observed and simulated variations, or simply the integrals of large scale quantities.

Most of the above mentioned metrics are useful for defining the overall model skill in simulating fields of common practical interest in the present climate, such as the surface air temperature or the accumulated precipitation. However, the fields in questions concern quantities that can hardly be considered *climate state variables*. By considering the vertical profile of the annual and global mean temperature (TAR, figure 8.8), the zonal mean surface air temperature (TAR, figure 8.2) or precipitation (TAR, figure 8.3), the impression is that all models have very similar performances and it is very difficult to assess whether a model is performing in any sense better than any other. Nevertheless, they differ substantially in the horizontal as well as vertical resolution, numerical schemes, physical parameterizations and so on.

If, instead of immediately checking how *realistic* – from the final user's point of view - the modelled fields of practical interest are, the aim is to plan strategies for model improvement, it is important to fully understand the differences in the representation of the *climatic machine* among models and possibly decide whether specific physical processes - typically related to energy/mass transport and energy conversion - are correctly simulated by a specific numerical model. In this context, a relevant example of a soundly-based studies are the atmospheric GCMs intercomparison on the representation of the hydrologic cycle performed by Lau et al. (1996) and the comparison of observed and simulated processes for the generation of atmospheric available potential energy (Siegmund, 1995).

The comparison of bulk thermodynamic quantities defining the climate state, such as the tropospheric average temperature, tropospheric average specific humidity, variance of geopotential height, allows the definition of *global metrics* which may be considered as robust diagnostic tools. Nevertheless, such approach does not allow for the disentanglement of the role of each one of the vast range of distinct physical processes contributing to the global balances.

In order to capture the differences in the representation of specific physical processes, it is necessary to use specialized diagnostic tools - that we may call *process-oriented metrics* - as indexes for model reliability. Such approach may be helpful in clarifying the distinction between the performance of the models in reproducing *diagnostic* and *prognostic* variables of the atmospheric system. Moreover, such approach would allow to highlight discrepancies in the statistical properties of the various terms contributing to the derivatives of the fields.

The 500hPa geopotential-height has an immediate application as a visualization tool for the organization of large scale winds, which are basically in geostrophic balance. However, it carries also important dynamical information. Theoretical as well as observational arguments suggest that the 500hPa geopotential-height can be taken as a key variable for describing atmospheric process both in the Low-Frequency Low-Wavenumber (LFLW) and in the High-Frequency High-Wavenumber (HFHW) region of the full spectrum of variability. HFHW phenomena comprise the synoptic travelling waves characterized by period of the order 2-7 days, by spatial scales of the order of few thousands kilometres, and which can be associated with the release of available energy driven by conventional baroclinic conversion (Blackmon 1976; Speranza 1983; Wallace et al. 1988), so that they are often referred to as baroclinic waves. LFLW phenomena comprise the lower frequency variability (period of the order of 10-45 days), mostly due to the dynamics of long stationary waves, interacting with orography (Charney and DeVore 1979; Charney and Straus 1980; Buzzi et al. 1984; Benzi et al. 1986) and being catalyzed by the sub-tropical jet (Benzi and Speranza, 1989; Ruti et al., 2006). Both the baroclinic and planetary waves provide a relevant contribution to the meridional transport of energy and momentum (Speranza, 1983; Peixoto and Oort, 1992). In a previous study (Dell'Aquila et al., 2005) we have compared the mid-latitude atmospheric variability in the northern hemisphere as described in two different reanalysis products. By performing a space-time spectral analysis (Hayashi, 1971, 1979) of the 500hPa geopotential height variability we were able to identify slight discrepancies in the description of the standing and propagating components of the total wave spectrum, due to differences in the observational basis as well as in the operational model formulations.

The operational global weather forecasting models routinely used in the '80s, which constitute in many cases the baselines of the atmospheric components presently adopted in up-to-date global climate models (for a brief history of atmospheric modelling, see http://www.aip.org/history/sloan/gcm/intro.html), were well-known to suffer from serious biases both in the LFLW and HFHW atmospheric variability domains (Tibaldi, 1986). In particular, it was confirmed in a number of studies that such models on the average featured a large overestimation of the baroclinic waves (e.g. Sumi and Kanamitsu, 1984; Klinker and Capaldo., 1986, Siegmund 1995) and a large underestimation of the planetary waves (e.g. Wallace et al., 1983), with biases sometimes of the same size as the average climatology of the observations and analyzed atmospheric fields. Similar biases have been identified also on climate models such as ECHAM 3 (Kaurola, 1996). Whereas such biases have a relatively minor impact on reanalyses, since observation are used to correct the autonomous evolution of the atmospheric model, they may be critical on unconstrained models, and may hinder the significance of the resulting simulations of past and future climatic conditions, with special regard to the mid-latitudes.

In view of the above mentioned remarks, in this study we consider 17 global climate models that will contribute to the scientific basis of the Intergovernmental Panel on Climate Change 4$^{th}$ Assessment Report (henceforth, IPCC-4AR) and perform an intercomparison study focusing on the mid-latitudes baroclinic and planetary waves, Moreover, we compare the statistics of the climate models with those of the NCEP-NCAR and ECMWF global reanalyses. To accomplish such a goal, we construct two different kinds of model metrics. The scope is to supplement the use of *global metrics,* describing the spectral properties of all sorts of atmospheric disturbances, which may average out the compensating effect of different physical processes, with a *process-oriented metrics* whose goal is to assess the model capability of correctly describing in detail a given climatic processes. Therefore, we introduce an *ad hoc* integral measure of the variability observed in the Northern Hemisphere on different spectral sub-domains. The total wave variability is taken as a global scalar metrics describing the overall performance of each model, while the total variability pertaining to the eastward propagating baroclinic waves and to the planetary waves are taken as scalar metrics describing the performance of each model in describing the HFHW variability and the LFLW variability, respectively. We maintain that such tests critically address the reliability of the considered models in the simulation of structurally major climatic processes.

This paper is organized as follows. In section 2 we describe the considered datasets and sketch the method employed for analysis. In section 3 we compare the model performances in specific sub-regions of the full space-temporal spectrum; the conclusions are drawn in section 4.

## 2. Data and methods

The IPCC models considered in this study are listed in table 1 along with their main features. Although the analysis is performed on an atmospheric variable, we report also a few details about the ocean component in order to stress that the results are representative of the response of a complex system (say non-linear) which can not be reduced to the behaviour of its single components (Lucarini, 2002). Indeed, in some cases (for example, the GISS model), the same atmosphere over different oceans results into substantially different atmospheric variability. The time frame considered is 1962-2000. For this period a control simulation has been performed by all IPCC models, imposing the observed concentration of trace gases, such as $CO_2$ and ozone.

The IPCC models output is compared with observations by considering two major reanalysis datasets: the reanalysis produced by the National Center for Environmental Prediction (NCEP), in collaboration with the National Center for Atmospheric Research (NCAR) (Kistler et al. 2001), and one released by the European Center for Mid-Range Weather Forecast (ECMWF) (Simmons and Gibson 2000). The resulting common period for all datasets is 1962-2000, which then results to be the time-frame we focus on in this work.

## *2.1 Geostrophic Approximation and Latitudinal Averaging*

Since the goal of this study is to diagnose the mid-latitude winter atmospheric variability of the considered climate models, along the lines of Dell'Aquila et al. (2005), we should use the December-January-February (DJF) daily values of geopotential height at 500hPa data averaged over the latitudinal belt 30°N-75°N, where the bulk of mid-latitude atmospheric waves activity is observed (Blackmon, 1976; Speranza, 1983).

Unfortunately, this field is not one of the standard daily 3D field outputs for the IPCC models, which comprises only zonal and meridional wind speed, air temperature and specific humidity. In principle, the geopotential height could be rigorously computed from temperature and specific humidity using the hydrostatic relations and the equation of state for air. Such an approach requires, however, knowledge of two time varying 3D fields (temperature and specific humidity) and of boundary terms such as the constant 2D field of surface height and the time-varying 2D field of surface pressure. The latter field is not readily available in the IPCC-4AR dataset, but could be reconstructed with suitable interpolations from the available sea-level pressure data.

In view of the large computational resources needed and the uncertainties on the surface boundary term, we have instead retrieved the 500 hPa meridional wind speed. In fact, in the geostrophic approximation, the meridional wind is related to the zonal gradient of the geopotential height by:

$$(1) \quad f(\phi) v(p, \phi, \lambda, t) = \frac{g}{R \cos(\phi)} \frac{\partial z(p, \phi, \lambda, t)}{\partial \lambda}$$

where $t$ is time, $p$, $\phi$, and $\lambda$ are respectively the pressure, the latitude and the longitude, $f(\phi) = 2\Omega \sin(\phi)$, $v$ indicates the meridional component of the wind velocity vector, $g$ is the gravity acceleration, $R$ is the radius of the Earth, and $z$ is the geopotential height. This approach requires much less computer resources, has a simple model-independent implementation, and is *local* in the sense that is involves only the relevant $p=500$hPa level. Of course the geostrophic relation is only an approximation at order Rossby number (about 0.05-0.10 in our case) approximation, but is well suited for the mid-latitudes (Peixoto and Oort, 1992; Holton, 1992), where we restrict our analysis. Moreover, the geostrophic relations provide the dynamical content of the geopotential height field: this is just the reason why the geopotential height is chosen in this and previous spectral studies of the mid-latitude atmospheric variability, so that such an approach is consistent with the goal of this study.

We can then obtain the following relation between the longitudinal derivative of the zonally averaged geopotential height and the meridional velocity (Peixoto and Oort, 1992):

$$(2) \quad \frac{\partial \langle z(p,\lambda,t)\rangle_{\phi_1}^{\phi_2}}{\partial \lambda} = \frac{R}{g} \langle f(\phi) v(p,\phi,\lambda,t) \cos(\phi) \rangle_{\phi_1}^{\phi_2}$$

where the area-weighted latitudinal average of the generic function $F(p,\phi,\lambda,t)$ is defined as follows:

$$(3) \quad \langle F(p,\phi,\lambda,t)\rangle_{\phi_1}^{\phi_2} = \frac{\int_{\phi_1}^{\phi_2} F(p,\phi,\lambda,t)\cos(\phi)d\phi}{\int_{\phi_1}^{\phi_2} \cos(\phi)d\phi}$$

We rename the function $\langle z(p,\lambda,t)\rangle_{\phi_1}^{\phi_2}$ as $Z(p,\lambda,t)$ and the function $R/g \langle f(\phi)v(p,\phi,\lambda,t)\cos(\phi)\rangle_{\phi_1}^{\phi_2}$ as $FV(p,\lambda,t)$. We then have:

$$(4) \quad \frac{\partial Z(p,\boldsymbol{l},t)}{\partial \boldsymbol{l}} = FV(p,\boldsymbol{l},t),$$

which constitutes the baseline of the later described spectral approach.

## *2.2 Hayashi spectra*

The variability of the one dimensional field in terms of waves of different periods and zonal wavenumbers can be effectively described by means of the space-time Fourier decomposition introduced by Hayashi (1971, 1979). By computing the cross-spectra and the coherence of the signal, the method allows for a separation of propagating and the standing components of the atmospheric waves.

Straightforward space-time decomposition will not distinguish between standing and travelling waves: a standing wave will give two spectral peaks corresponding to travelling waves moving eastward and westward at the same speed and with the same phase. The problem can only be circumvented by making assumptions regarding the nature of the wave. One approach relies in attributing complete coherence between the eastward and westward components of standing waves and on attributing the incoherent part of the spectra to real travelling waves (Pratt, 1976, Fraedrich and Bottger, 1978; Hayashi,1979).

In this formulation, for each winter considered, it is possible to express the quantity $X(\boldsymbol{l},t)$ in terms of the its zonal Fourier harmonic components as:

$$(5) \quad X(\boldsymbol{l},t) = X_0(t) + \sum_{j=1}^{\infty} \left\{ C_{k_j}(t)\cos(k_j \boldsymbol{l}) + S_{k_j}(t)\sin(k_j \boldsymbol{l}) \right\}$$

where t ranges between 0 and the winter length $t = 90d$, the zonal wavenumber is expressed as $k_j = j$ and $\boldsymbol{l}$ ranges between 0 and $2\boldsymbol{p}$.

The power spectrum $H_{E/W}(k_j,\boldsymbol{w}_m)$ at a zonal wavenumber $k_j$ and temporal frequency $\boldsymbol{w}_m = 2\boldsymbol{p}m/t$ for the eastward and westward propagating waves is:

$$(6) \quad H_E(k_j, w_m) = \frac{1}{4}\{P_{w_m}(C_{k_j}) + P_{w_m}(S_{k_j})\} + \frac{1}{2}Q_{w_m}(C_{k_j}, S_{k_j})$$

$$(7) \quad H_W(k_j, w_m) = \frac{1}{4}\{P_{w_m}(C_{k_j}) + P_{w_m}(S_{k_j})\} - \frac{1}{2}Q_{w_m}(C_{k_j}, S_{k_j})$$

where $P_{w_m}$ and $Q_{w_m}$ are, respectively the power and the quadrature spectra of zonal Fourier harmonic of $X(l,t)$.

The total variance spectrum $H_T(k_j, w_m)$ is given from the sum of the eastward and westward propagating components:

$$(8) \quad H_T(k_j, w_m) = \frac{1}{2}\{P_{w_m}(C_{k_j}) + P_{w_m}(S_{k_j})\}$$

while the propagating variance $H_P(k, w)$ is given by the difference between the components (A1a) and (A1b):

$$(9) \quad H_P(k_j, w_m) = |Q(k_j, w_m)|.$$

So, the standing variance spectrum $H_S(k, w)$ can be obtained by the difference:

$$(10) \quad H_S(k_j, w_m) = H_T(k_j, w_m) - |Q(k_j, w_m)|.$$

We emphasize that for sake of simplicity of the notation, we have neglected the indication of the winter under investigation, denoted in the text by the superscript $n$. We emphasize that, customarily, Hayashi spectra are generally represented by plotting the quantities $j \cdot m \cdot H_T(k_j, w_m)$, $j \cdot m \cdot H_S(k_j, w_m)$, $j \cdot m \cdot H_E(k_j, w_m)$, and $j \cdot m \cdot H_W(k_j, w_m)$, in order for equal geometrical areas in the log-log plot to represent equal variance.

By considering the equations (5-10) descriptive of the Hayashi spectra and the basic properties of transformation of the Fourier series with respect to the derivation, we have that the following relations hold between the Hayashi spectra of the function $Z(p,l,t)$ and of the function $FV(p,l,t)$:

$$(11) \quad H_a(k_j, \mathbf{w}_m)\big|_Z = k_j^2 H_a(k_j, \mathbf{w}_m)\big|_{FV}$$

where the subscript $a$ can take the values of $T, S, E, W$, which refer to the total variance and to standing eastward propagating, and westward propagating components of the spectrum, respectively.

Therefore, in this work, for each considered dataset, we first compute the Hayashi space-time spectra of the quantity $FV(p, \mathbf{l}, t) = R/g \langle f(\mathbf{f}) v(p, \mathbf{f}, \mathbf{l}, t) \cos(\mathbf{f}) \rangle_{\mathbf{f}_1}^{\mathbf{f}_2}$ with $p = 500 hPa$, $\mathbf{f}_1$ and $\mathbf{f}_2$ set to the grid-points closest to 30°N and 75°N, respectively, and then obtain the Hayashi spectra of the corresponding latitudinally averaged geopotential height by using equation (11).

In order to evaluate the model performances in different spectral sub domains, we introduce the following integral quantities:

$$(12) \quad E_a^n(\Omega) = \sum_{m=m_1, j=j_1}^{m=m_2, j=j_2} H_a^n(k_j, \mathbf{w}_m),$$

where $a=T,S,E,W$, $n$ indicates the winter; the integration extremes, $m_{1,2}$ and $j_{1,2}$, determine the spectral region of interest $\Omega = [\mathbf{w}_{m_1}, \mathbf{w}_{m_2}] \times [k_{j_1}, k_{j_2}]$. The quantity $E_a^n(\Omega)$ introduced in equation (6) represents the portion of variance of the spectrum associated to a given subdomain $\Omega$ and to a given winter $n$ and is expressed in units of $m^2$. The averaging process defined in equation (12) overcomes the well-known instability of the Fourier analysis in describing small scale spectral features. Following basic statistical arguments, we estimate the average intra-seasonal atmospheric variability in the spectral subdomain $\Omega$ as:

$$(13) \quad \overline{E}_a(\Omega) = \frac{1}{N} \sum_{n=1}^{N_2} E_a^n(\Omega),$$

where $N$ is the number of years considered in the averaging process. The interannual variability of the signal $E_t^n(\Omega)$ is described in terms of its standard deviation:

$$(14) \quad \boldsymbol{s}_{E_a(\Omega)} = \sqrt{\frac{1}{N-1}\sum_{n=1}^{N}\left(E_a^n(\Omega) - \overline{E}_a(\Omega)\right)^2}.$$

The two quantities $\overline{E}_a(\Omega)$ and $\boldsymbol{s}_{E_a(\Omega)}$ characterize the climatology of the atmospheric waves occurring in the spectral subdomain $\Omega$.

It is possible to test the reliability of the Hayashi spectra of the reconstructed geopotential height and to estimate the essentially model-independent biases introduced by the geostrophic approximation by including in this study the analysis of the variability of the readily available 500hPa geopotential height of the NCEP-NCAR and ECMWF reanalyses, thoroughly studied in Dell'Aquila et al. (2005).

## 3. Results

The space-time spectra is computed for each of the 39 winters included in our datasets. Fig. 1a-d show the various components of the 39-winters averages of the spectra as computed from the *geostrophically reconstructed* 500 hPa geopotential height for the NCEP reanalysis dataset averaged over the latitudinal band 30°N-75°N. Fig. 1a shows the total power spectrum; Fig. 1b shows the power spectrum related to standing waves; Fig. 1c shows the power spectrum related to eastward propagating waves; Fig. 1d shows the spectrum of the westward propagating waves.

For comparison, we show in Fig. 2 the Hayashi spectra computed with the *actual* 500 hPa geopotential height for the NCEP reanalysis dataset averaged over the latitudinal band 30°N-75°N, already shown and discussed in Dell'Aquila et al. (2005). Visual inspection shows that the results are rather similar to what presented with the geostrophically reconstructed field. Nevertheless, we notice that the Hayashi spectra of the actual 500hPa (Fig. 2) have consistently slightly higher values (about 10%) than those computed with the reconstructed field, thus suggesting that the geostrophic reconstruction filters out some atmospheric variability, as to be reasonably expected. We obtain similar results by reconstructing geopotential height field from ERA40 reanalysis (not shown).

Since the effects of geostrophic approximation can be safely considered as essentially model-independent, we can then assume the correctness of our approach and consider the Hayashi spectra of the latitudinally averaged and geostrophically reconstructed 500 hPa geopotential height fields of the various models as good proxies for the spectra of the exact fields.

The Hayashi spectra of the considered IPCC models reported in Table 1 are presented in Figs. 3-6. All models spectra are, after visual inspection, qualitatively similar to those of the NCEP and ERA40 datasets. In particular, a large portion of the total variance is concentrated in the low frequency – low wavenumber domain, and can be related mostly to standing waves and to westward propagating waves. The high frequency – high wavenumber domain, corresponding mainly to synoptic disturbances, contains a smaller portion of the total variance, and is related to eastward propagating waves.

Nevertheless, it is clear that many models seriously underestimate the power spectrum of the atmospheric signal at most time and space scales, such as the CGCM3.1 (T42), the CNRM-CM3, the CSIRO-Mk3.0, the ECHO-G, the entire family of GISS GCMs, the MRI-CGCM2.3.2. It is clear that such models provide a particularly serious underestimate for the standing and westward propagating waves. On the contrary, some other models, such as ECHAM5/MPI-OM and FGOALS-g1.0 tend to overestimate the atmospheric variability at all scales, and provide a more extreme overestimate of the eastward propagating waves.

Another serious issue concerning the performances of the atmospheric components of climate models is their ability of describing the statistic of standing waves. In fact, planetary standing waves are a major feature of mid-latitude atmospheric dynamics in the northern hemisphere, where ultra-long waves may resonate with topography (Charney and DeVore 1979). However, this feature appears to be very hard to be captured by most of the models considered here. As shown in Fig. 5, a well defined spectral peak is present at wave number 3 and period of about 20 days for both reanalyses. Instead, most of the considered models exhibit either very broad peaks in the low frequency-low wavenumber region of the full spectrum (CGCM3.1(T42), CGCM3.1(T63), CNRM, ECHO-G, FGOALS, GFDL-CM2.1, GISS-ER), or the peak is in the wrong position (GFDL-CM2.0, GISS-EH, INM-CM3.0), or multiple peaks exist (CSIRO-Mk3.0, ECHAM5/MPI-OM). The only examples of models that correctly describe this aspect of mid-latitude atmospheric variability are the IPSL-CM4 and MIROC models.

It is clear that visual inspection of the Hayashi spectra, though instructive, is not a very efficient way for objectively evaluating the model's performances. We then try to summarize the most relevant information contained in the panels showing the Hayashi spectra into a few numbers that can be more easily employed to characterize the skill of each model. We next compare to different approaches by measuring the overall performance of each models at all time and space scales (*global metrics*) and the ability of the models in reproducing the correct features of the baroclinic and planetary waves (*process oriented metrics*).

## *3.1 Global metrics*

A global scalar metrics can be introduced by integrating the whole space-time power spectrum corresponding to each one of the 39 considered winter seasons. We then consider the quantity $E_T^n(\Omega)$ introduced in equation (12) where O is set to be the full frequency-wavenumber domain and *n* is the index running over the considered winters.

In Fig. **7** we show for all the considered datasets the climatological average $\overline{E}_T(\Omega)$ of the integral of the full spectrum versus its interannual variability $s_{E_T(\Omega)}$, estimated according to equations (13) and (14), respectively. In this scatter diagram the abscissas represent the model average intra-seasonal variability, while the ordinates represent the interannual variability of the model intra-seasonal variability. The models ensemble average is also shown. Such scatter diagram allows a complete visual representation of a given climate by reporting the two most relevant statistical moments of any chosen variable.

The *reconstructed* latitudinally averaged 500 hPa geopotential height fields of the two reanalyses have similar intra-seasonal variability and slightly different interannual variability mainly because of the discrepancies in the pre-satellite period. In fact, it has been shown in Dell'Aquila et al. (2005) that in the pre-satellite period a bias between the two reanalyses exists, with ERA40 featuring systematically larger variability of the high frequency-high wavenumber eastward propagating waves. If we consider the *observed* fields (here not shown), in agreement with what hinted from the visual inspection of the Hayashi spectra, we have that in both cases the intraseasonal and interannual atmospheric variability is increased by a constant factor around 10%, which proves that the geostrophic approximations acts as a filter having an overall gain smaller than 1. The robustness of our approach is then confirmed.

Apparently, if we consider the model ensemble, we have a good agreement with observations. Both reanalyses lie well within one standard deviation from the ensemble average (center of the ellipses in Fig. **7**). However, the models are widely spread over the plane space spanned by the two variables, with a typical bias of about 15% with respect to the reanalyses. In particular, considering the biases on the quantity $\overline{E}_T(\Omega)$, and considering that the standard deviation of the climatological mean $\overline{E}_T(\Omega)$ can be approximated as $s_{E_T(\Omega)}/\sqrt{N} \approx 0.16 s_{E_T(\Omega)}$, since the winters are weakly dependent, we have that very few models are statistically consistent with the reanalyses with a reasonable significance.

In this sense, the best models are by far the high-resolution version of the MIROC and the GFDL-CM2.1. The T62 version of the CGCM3.2 and the relatively low-resolution INM3.0 also perform well, featuring a slightly too large intraseasonal variability and too low interannual variability, respectively. Some models, such FGOALS1.0 and ECHAM5/MPI-OM, feature a very large positive biases ranging over 20% for the interannual and intraseasonal variability. Some other models, usually of relatively low-resolution, such as CNRM-CM3 GISS-ER, ECHO-G and MRI-CGCM2.3.2, feature over 20% negative biases. In general, the biases on the intraseasonal and interannual variability are positively linearly correlated: for larger average signals the variability tend to be larger, so that the model spread in Fig. **7** is definitely not isotropic.

In some cases, it is possible to track the improvements occurring between different versions or set-ups of the same climate model. Thus, *e.g.*, the GISS-EH model (which includes an isopycnal ocean component) has a better representation of the intra-seasonal atmospheric with respect to the GISS-ER which has a z-coordinate ocean model. However, the two models have considerably different interannual variability, which is in both cases different from that of the reanalysis. The CGCM3.1 model is presented in two versions which are identical except for the horizontal resolution (T47, corresponding to about 3.75° resolution, and T63 corresponding to about 2.8° resolution). The CGCM3.1(T63) improves the representation of both the intra-seasonal variability and of the interannual variability with respect to CGCM3.1(T47). Similar improvements are observed between MIROC(hires) and MIROC(medres) and between GFDL-CM2.1 and GFDL-CM2.0. The MIROC model is presented with two different horizontal as well as vertical resolution (T42L20 for the *medres* version and T106L56 for the *hires* version). The *medres* version shows already quite good performances and is among the best models. However, a substantial improvement is observed when switching to higher resolution. The GFDL-CM2.0 and GFDL-CM2.1 models are very similar. In particular they share the same horizontal as well as vertical resolution, but in the GFDL-CM2.1 model some numerical techniques are improved with respect to GFDL-CM2.0. For example, the horizontal in CM2.0 uses centered spatial differencing, whereas in CM2.1 the horizontal discretization is performed with a flux-form semi-Lagrangian method. Finally, let us note that the models with a better representation of intraseasonal and interannual variability display also the most realistic ENSO variability as shown in Van Oldenburg et al. (2005).

### *3.2 Process-oriented metrics*

In order to construct a *process-oriented* metrics pointing at the diagnostics of specific physical processes, we consider a suitable decomposition of the whole frequency-wave number domain. Following Dell'Aquila et al. (2005), we consider two spectral subdomains: the LFLW subdomain, which includes periods from 10 to 45 days ($2 \leq m \leq 9$) and zonal wavenumbers $2 \leq j \leq 4$ (length scales larger than 7000Km); the HFHW subdomain, which includes periods from 2 to 7 days ($13 \leq m \leq 36$) and zonal wavenumbers $j \geq 6$ (length scales ranging from a few hundreds of kilometres to 5000Km). For each year of a given dataset, we then provide a bulk measure of the planetary standing waves and of the eastward propagating baroclinic waves by $E_S^n(\Omega_{LFLW})$ and $E_E^n(\Omega_{HFHW})$ defined in equation (12), respectively, where the two O-domains are prescribed as above, *n* is the index of the winters, and the lower indexes *S* and *E* refer to standing and eastward propagating components, respectively. The quantities $E_S^n(\Omega_{LFLW})$ and $E_E^n(\Omega_{HFHW})$ can then be used to characterize the capabilities of each model of correctly describing two specific physical processes of the atmosphere having wave nature.

We focus on the average (climatological) description provided by each model of the planetary standing waves and of the baroclinic eastward propagating waves and keep slightly aside the problem of checking the correctness in the description of the interannual variability of the two signals. We then consider for each model the quantities $\overline{E}_S(\Omega_{LFLW})$ and $\overline{E}_E(\Omega_{HFHW})$ with the corresponding standard error estimated as expressed as $s_{E_S(\Omega_{LFLW})}/\sqrt{N}$ and $s_{E_E(\Omega_{HFHW})}/\sqrt{N}$, respectively. This entails assuming a weak time-lagged correlation between the 39 winter seasons for the two signals in each model, which is correct in the first approximation.

In Fig. 8 we show for all the considered datasets the climatological average $\overline{E}_S(\Omega_{LFLW})$ versus the climatological average $\overline{E}_E(\Omega_{HFHW})$, and indicate for both directions the corresponding standard error. When considering the ERA40 and NCEP renalyses datasets, we observe that the two reanalyses are very close to each other and they could be brought to a closer agreement by dropping the pre-satellite period (Dell'Aquila et al., 2005). However, the discrepancies among IPCC models by far exceed the discrepancies between the two different *dynamical interpolations* of the available observations.

In particular, with this *process oriented* metrics, the agreement with observations seems weaker than with the *global* metrics, with the two reanalysis lying at just about one standard deviation from the ensemble average. It is then important to stress that with the *global* metrics, the errors originating in different regions of the spectrum may average out and hide major model deficiencies. Instead, in with the *process oriented* metrics, it is easier to bring the model's biases into light.

For example, a general (with very few exceptions) occurrence for all IPCC models is that they overestimate the variability in the HFHW subdomain. Regarding the LFLW subdomain, more than half of the models tend to underestimate the corresponding variability. As an overall result, the model ensemble (center of the ellipses in Fig. 8) overestimates the HFHW variability and underestimates the LFLW variability. This closely resembles the results of many diagnostic studies performed in the past on global weather forecasting models (Tibaldi, 1986), which constitute the ancestors of the atmospheric components of the considered climate models.

Moreover, if we consider the standard deviation of the climatological means we again have that very few models are statistically consistent with the reanalyses with a reasonable significance. The models that are closer to the reanalyses are the MIROC(hires) model and the GFDL2.1 model. In these two cases, the error bars have some overlap with the area defined by the error bars of the NCEP and ERA40 reanalyses. By using this metrics, the models that appear to have the worst performances are the CNRM-CM3 model, which underestimates both the low frequency and the high frequency, and the FGOALS model, which overestimates the variability in both spectral sub-domains.

Unfortunately, there is no unique way to attribute the biases of LFLW and HFHW variability to specific features that are common to a class of models. Instead, there are cases in which different versions of the same model show different behaviour in this spectral subdomain. We refer to the CGCM3.1-T47 and CGCM3.1-T63 models, which differ for the horizontal resolution, and the GISS-EH and GISS-ER models, which share the same atmospheric component over different ocean models. Therefore, it is interesting to compare the relative changes in the performances of models presented with different configuration. This might help in the identification of specific strategies for model improvements.

In the case of the CGCM3.1 models, an increased horizontal resolution (T63) leads to a better statistics of planetary-scale standing waves. However, the performance on small scale disturbances is better in the low resolution version (T47). Thus, increasing the horizontal resolution alone leads to no automatic overall substantial improvement of the model performance.

As mentioned above, the two versions of the GISS model share the same atmosphere component over different oceans. In this case, the use of vertical density coordinates (GISS-EH) appears to improve the statistics of the standing waves which is too low in the case of z-coordinates (GISS-ER). Instead, the performance in on eastward propagating waves is worse.

In the case of the MIROC model, increased resolution alone improves the model performances. However, unlike with the CGCM3.1 models, in this case both the horizontal *and* the vertical resolution is increased.

Also in the context of process-oriented metrics, the improvement from GFDL-CM2.0 to GFDL-CM2.1 is substantial, with the latter in good agreement with the reanalyses. The experience with the GFDL models demonstrates how the increase of computer power (*i.e.* of resolution) may not be the only pathway to model improvements. Instead, the study of more accurate discretization techniques may give a substantial contribution.

## 4. Conclusions

The goal of this study is the evaluation of the degree of realism and of mutual coherence of some of the most well-known GCMs in the description of statistical properties of the atmospheric disturbances in the free atmosphere in the present climate. We maintain that such analysis is more insightful into the real performances of the GCMs than the comparison of essentially boundary properties such as surface temperature, because the internal mechanisms of the atmosphere are here taken into consideration.

We have considered the variability of the 500hPa geopotential height field, as described in the NCEP and ERA40 reanalyses for the time frame 1962-2000 and in the XX century control run of the IPCC GCMs. We compute the Hayashi spectra of the 500hPa geopotential height fields and introduce an *ad hoc* integral measure of the variability observed in the Northern Hemisphere on different spectral sub-domains. The total wave variability is taken as a *global metrics* describing the overall performance of each model, while the total variability pertaining to the eastward propagating baroclinic waves and to the standing planetary waves, respectively are taken as process-oriented metrics, aimed at measuring the model capability of describing the corresponding physical process.

In such a context, we obtain the striking result that large biases, in most cases larger than 10%, are found in all the considered metrics between the atmospheric waves climatology of most IPCC models and the reanalyses. The span of the climatologies of the various models is in all cases around 50% of the climatology of the reanalyses. In particular, when considering the total variability of the wave fields of the GCMs, we have that the biases on the intraseasonal and interannual variability are positively linearly correlated: for larger average signals the variability tend to be larger. When considering the process-oriented metrics, we have that the baroclinic waves are typically overestimated by the climate models, while the planetary waves are usually underestimated. This closely resembles the results of many diagnostic studies performed in the past on global weather forecasting models (Tibaldi, 1986). The climatologies of the wave activity of only two models – GFDL-CM2.1 and MIROC(hires) - are statistically consistent with that of the reanalyses both for the global and process-oriented metrics. This is a rather surprising result, given that all models are expected to provide very similar vertical temperature profiles, average surface temperature, precipitation and so on (see *e.g.* the results presented in the TAR).

Nevertheless, the general pictures obtained with the *global* and with the *process-oriented* metrics, are substantially different. In particular the apparent substantial improvement detected in the *global metrics* (as in the case of the CGCM3.1 model) may indeed mask the loss of performance in describing a specific process. Also, the INM-CM3.0 model, which seems rather close to observations with the *global metrics*, fails to describe correctly all regions of the spectrum of atmospheric variability at mid-latitudes. On the other hand, the *process-oriented metrics* may suggest some of the priorities for planning strategies for model improvements. In this perspective, we found that the increase of horizontal resolution alone has no substantial effect on model performances while the increase of horizontal *and* vertical resolution brings the MIROC(hires) model into close agreement with observations. An increased vertical resolution could be useful to better mimic the vertical structure of the ultra-long waves, in particular the orographic baroclinic standing perturbations (Buzzi et al. 1984). The improvement of numerical scheme has also a positive impact on model performances (GFDL models). In particular, the use of semi-lagrangian advection schemes for tracers seems to be an important requirement for model reliability. Somewhat unexpectedly, in the case of the GISS-ER and GISS-EH models, the characteristics of the adopted ocean model also seems to be critical in determining the agreement with the reanalyses. Among the three GISS models, the GISS-AOM seems to have superior performances. The models ensemble obtained by arithmetic averaging of the results of all models is biased with respect to the reanalyses but is comparable to the best 5 models.

This study suggests a serious caveat concerning the ability of most of the presently available climate models in describing the statistical properties of the global scale atmospheric dynamics of the present climate, and, *a fortiori*, in the perspective of climate change. One of the possible outcomes of this study may be the provision of quantitative information needed to *weight* model reliability when considering models ensemble results, *e.g.*, in the context of the IPCC reports.

On the other hand, the GFDL-CM2.1 and MIROC(hires) models, being able to reproduce with some degree of confidence the statistical properties of wave activity in the atmosphere, seem to be the best candidates for more detailed studies on atmospheric circulation regimes (Ruti et al., 2006), which will be the subject of future study. Among the several other in-depth analyses which can follow up from the results presented here, we would like to mention two future paths. In the context of the global properties of the atmosphere, it seems relevant to study the links between the degree of the models mutual coherence and realism in the description of the mid-latitudes atmospheric variability and in the representation of ENSO (Van Oldenburg et al. 2005), which seems critical in the set-up of the regimes of the low-frequency mid-latitudes planetary waves (Ruti el al., 2006). In the context of the understanding of climate change, it seems relevant to study the mutual coherence of the GCMs in their sensitivity to $CO_2$ doubling of the statistics of the atmospheric waves considered in the present analysis.

A complete and versatile toolbox written in MATLAB® language for the production of all figures and diagnostics contained in this paper is available upon request to the authors of this paper.

**Acknowledgements**

We acknowledge the international modelling groups for providing their data for analysis, the Program for Climate Model Diagnosis and Intercomparison (PCMDI) for collecting and archiving the model data, the JSC/CLIVAR Working Group on Coupled Modelling (WGCM) and their Coupled Model Intercomparison Project (CMIP) and Climate Simulation Panel for organizing the model data analysis activity, and the IPCC WG1 TSU for technical support. The IPCC Data Archive at Lawrence Livermore National Laboratory is supported by the Office of Science, U.S. Department of Energy. A special thank to George Boer, Seita Emori, Kenneth Lo, Seung-Ki Min, Daniel Robitaille and Gavin Schmidt for supplying some of the datasets that, due to technical problems, were not available on the on the PCMDI servers when we were working on this paper.

|  | Model (Reference) | Institution | Atmosphere | | | | Ocean | | |
|---|---|---|---|---|---|---|---|---|---|
|  |  |  | Trunc. (Lon x Lat) | Z | Chemistry and Aerosols | Albedo | Numerics and Adjustments | Lon x Lat. (Equator) | Vertical |
| 1. | CNRM-CM3 Salàs-Melia et al. (2005) | Mètèo France, *France* | T63 | h45 | $O_3$, $mA_c$, $dA_c$, $uA_p$, $sA_p$ | - | L, sI | 2° x 2° (2° x 0.5°) | 31z |
| 2. 3. | CGCM3(T47) CGCM3(T63) Kim et al. (2002) | CCCma, *Canada* | T47(T63) | z31 | - | - |  | 1.85° x 1.85° | 29z |
| 4. | CSIRO-Mk3.0 Gordon et al. (2002) | CSIRO, *Australia* | T63 | h18 | $sA_p$ | $L_p$ | L, Lp | 1.875° x 0.84° | 31z |
| 5. | ECHAM5/ MPI-OM Jungclaus et al. (2005) | Max Planck Inst., *Germany* | T63 | h31 | $sA_p$ | $I_m$, $L_m$ | L, sI/Lp | 1.5° x 1.5° | 40z |
| 6. | ECHO-G Min et al. (2005) | MIUB, METRI, and M&D *Germany/Korea* | T30 | h19 | $sA_p$ | $I_m$, $L_m$ | L, sI/Lp, F | 2.8° x 2.8° (0.5° x 0.5°) | 20z |
| 7. | FGOALS-g1.0 Yu et al. (2004) | LASG, *China* | 2.8° x 2.8° | σ26 | sA | - | O, sI, HWnC | 1° x 1° | H |
| 8. 9. | GFDL-CM2.0 GFDL-CM2.1 Delworth et al. (2005) | GFDL, *USA* | 2.5° x 2.0° | 24 | $mA_p$, $dA_c$, $uA_p$, $sA_p$ | - | C (CM2.0) L (CM2.1) | 1° x 1° (1° x 1/3°) | -z |
| 10. | GISS-AOM Lucarini and Russel (2002) | NASA-GISS, *USA* | 4° x 3° | h12 | $sA_p$ | $I_m$, $L_m$ | U, Lp | 4° x 3° | 16σ |
| 11. 12. | GISS-EH GISS-ER Schmidt et al. (2005) | NASA-GISS, *USA* | 4° x 5° | - | $mA_p$, $dA_p$, $uA_p$, $sA_p$ | $I_c$, $L_m$ | U | 2° x 2° (EH) 4° x 5° (ER) | σ (EH) z (ER) |
| 13. | INM-CM3.0 Volodin and Diansky (2004) | Inst. Of Num. Math., *Russia* | 5° x 4° | 21 | sA | $I_p$, $L_m$ | C, sI, F | 2.5° x 2° | 33σ |
| 14. | IPSL-CM4 Marti et al. (2005) | IPSL, *France* | 2.4° x 3.75° | 19 | - | $I_m$, $L_m$ | C, Lp | 2° x 2° (2° x 0.5°) | 31z |
| 15. 16. | MIROC3.2(*hires*) MIROC3.2(*medres*) K-1 mod. Dev. (2004) | CCSR/NIES/FRCGC, *Japan* | T106(*hires*) T42(*medres*) | 56(*hires*) 20(*medres*) | $mA_m$, $uA_m$, $sA_m$ | $I_m$, $L_m$ | L, Lp | 1.4° x 1.4° (1.4° x 0.5°) | 43h |
| 17. | MRI-CGCM2 Yukymoto and Noda (2002) | Meteorological Research Institute, *Japan* | T42 | 30 | sA | $I_m$, $L_m$ | S, Lp, F | 2.5° x 2.0° (2.5° x 0.5°) | ?z |

**Table 1.** An overview of the IPCC models. The horizontal resolution is expressed in terms of truncation (T) for spectral models (with T47 ≈ 2.8°, T63 ≈ 2.8°, T63 ≈ 2.8°); for Z coordinates, the letter before the number of levels indicates whether the vertical coordinate is height (z), pressure(p), pressure normalized with surface pressure (σ) or hybrid (h); in the chemistry column, the chemical species inculded in the model are specified; in the same column, the string $xA_y$ indicates whether the included aerosols are marine (x=m), desertic (x=d), urban (x=s) or sulfates (x=s) and if they are imposed as constants (y=c), prescribed according to climatolgy (y=p) or modeled (y=p); other model components may include biogeochemistry (B), vegetation (V), ice-sheets (I); Clouds ($C_y$) can be statistical (y=s) or modeled (y=m), the treatment of albedo is reported for oceanic albedo ($O_y$), ice albedo ($I_y$) and land surface albedo ($L_y$) that can be constant (y=c) or modeled (y=m); some details about the numerics are also specified, in particular it is specified whether the advection scheme for heat and moisture is or centered (C), upstream (U), spectral (S) or semi-Lagrangian (L); if the time stepping is explicit – Leapfrog (Lp) - or semi-implicit (sI) and whether flux-adjustment is employed (F). For the oceanic component we report the horizontal resolution (resolution at the equator in parenteses if it is different from the rest of the domain); the number of vertical levels along with an indication of the vertical coordinate that can be depth (z), depth normalized with the maximum local depth (σ), hybrid (h) or density (d). For further information, refer to the PCMDI web site http://www-pcmdi.llnl.gov.

| Spectral properties | $k_1 = 2, k_2 = 4$ | $k_1 = 6, k_2 = 72$ |
|---|---|---|
| $w_1 = (2p/45)d^{-1}, w_2 = (2p/10)d^{-1}$ | $\Omega$ = LFLW | $\Omega$ = LFHW |
| $w_1 = (2p/7)d^{-1}, w_2 = (2p/2)d^{-1}$ | $\Omega$ = HFLW | $\Omega$ = HFHW |

**Table 2:** Definition of 4 regions in the Hayashi spectra of the winter atmospheric variability; the symbol $d$ is used as shorthand for 'day'. *LFLW*: Low Frequency Long Wavenumber; *HFLW*: High Frequency Long Wavenumber; *LFHW*: Low Frequency High Wavenumber; *HFHW*: High Frequency High Wavenumber. Low Frequency relates to periods from 10 to 45 days; High Frequency relates to periods from 2 to 7 days; Low Wavenumber relates to length scales larger than 1000Km; High wavenumber relates to length scales ranging from a few tens to hundreds of kilometers. The values $w_2 = (2p/2)d^{-1}, k_2 = 72$ constitute the highest frequency and wavenumber allowed by the adopted data resolution.

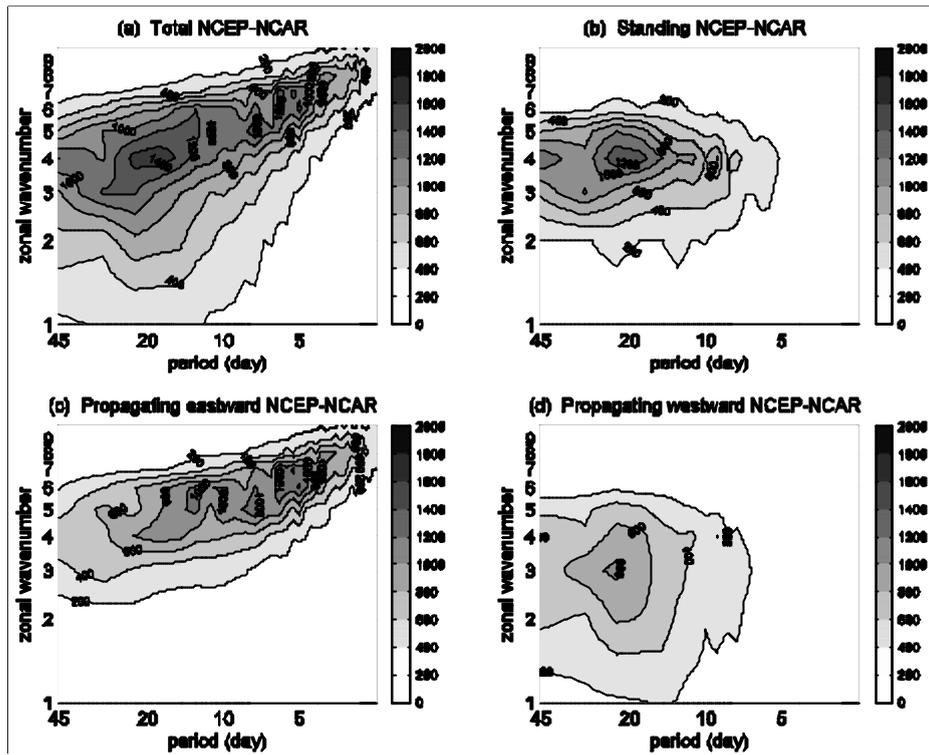

Figure 1: Climatological average over 39 winters of Hayashi spectra for the geostrophically reconstructed 500 hPa geopotential height (relative to the latitudinal belt 30°N-75°N) from NCEP data. Hayashi spectra are multiplied times $k\omega\tau/2\pi$ for representation purposes. The units are $m^2$.

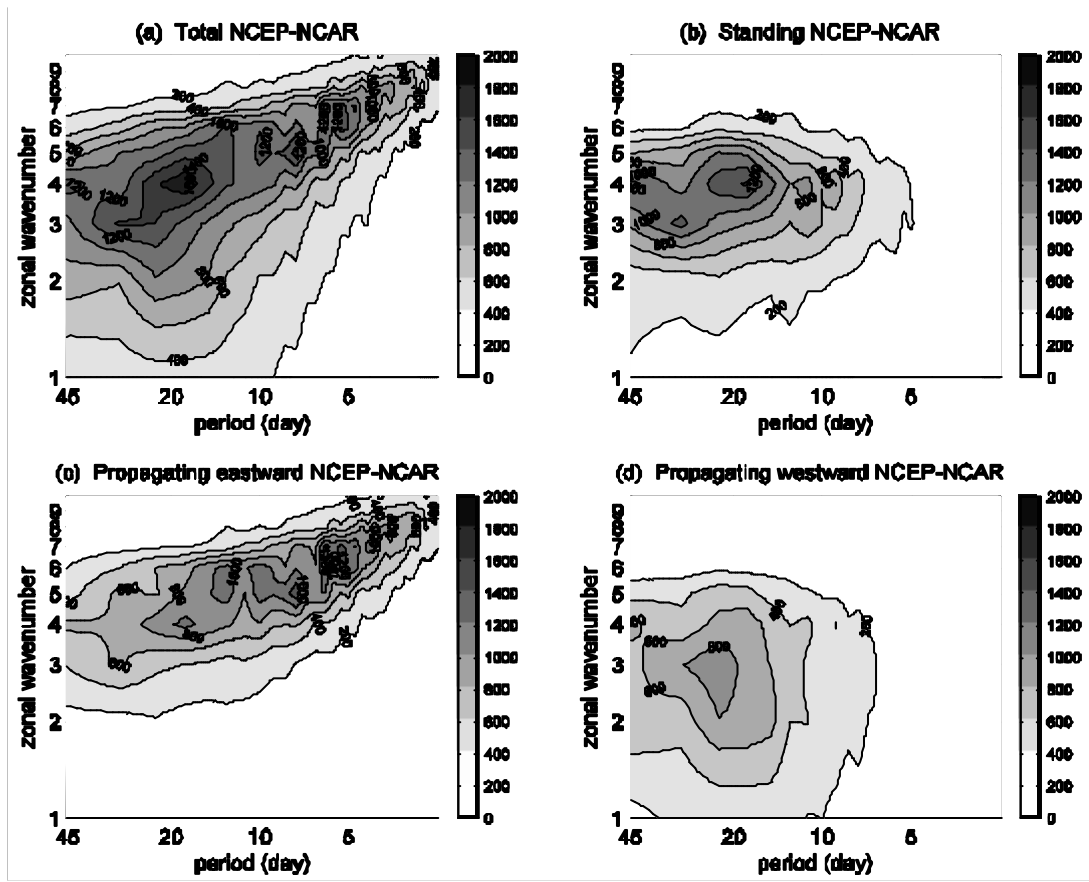

Figure 2: Climatological average over 39 winters of Hayashi spectra for the 30°N-75°N averaged 500 hPa geopotential height from NCEP data. Hayashi spectra are multiplied times $k\omega\tau/2\pi$ for representation purposes. The units are $m^2$.

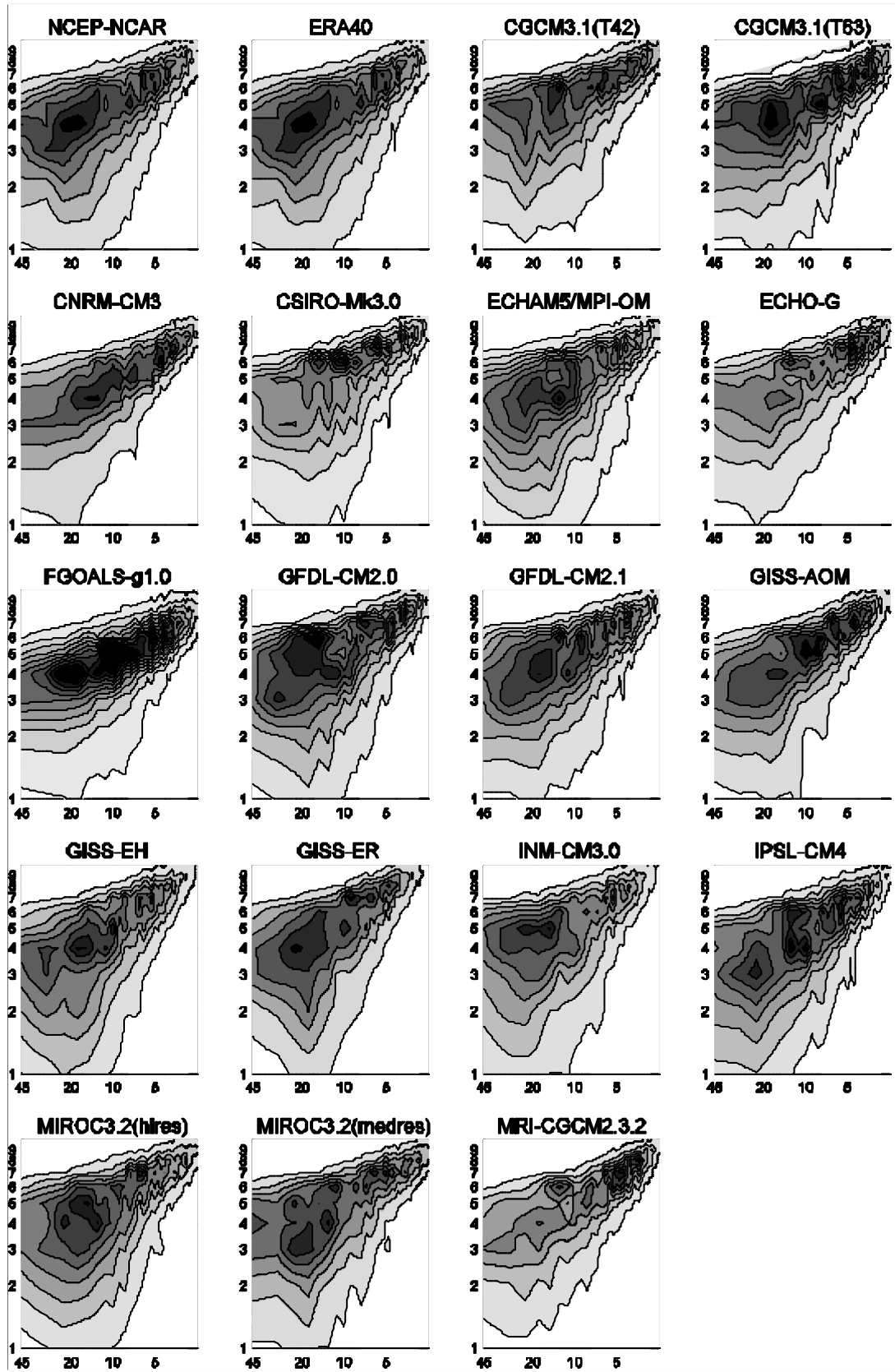

Figure 1: Climatological averages over 39 winters of the total Hayashi spectra for the geostrophically reconstructed 30°N-75°N averaged 500 hPa geopotential height from the 2 reanalyses and the 17 GCMs, as indicated in the panels. Hayashi spectra are multiplied times $k\omega\tau/2\pi$ for representation purposes (see text). Darker shades of grey denote higher values and isolines intervals are 200 $m^2$. Period (in days) in abscissas and wavenumber in ordinates.

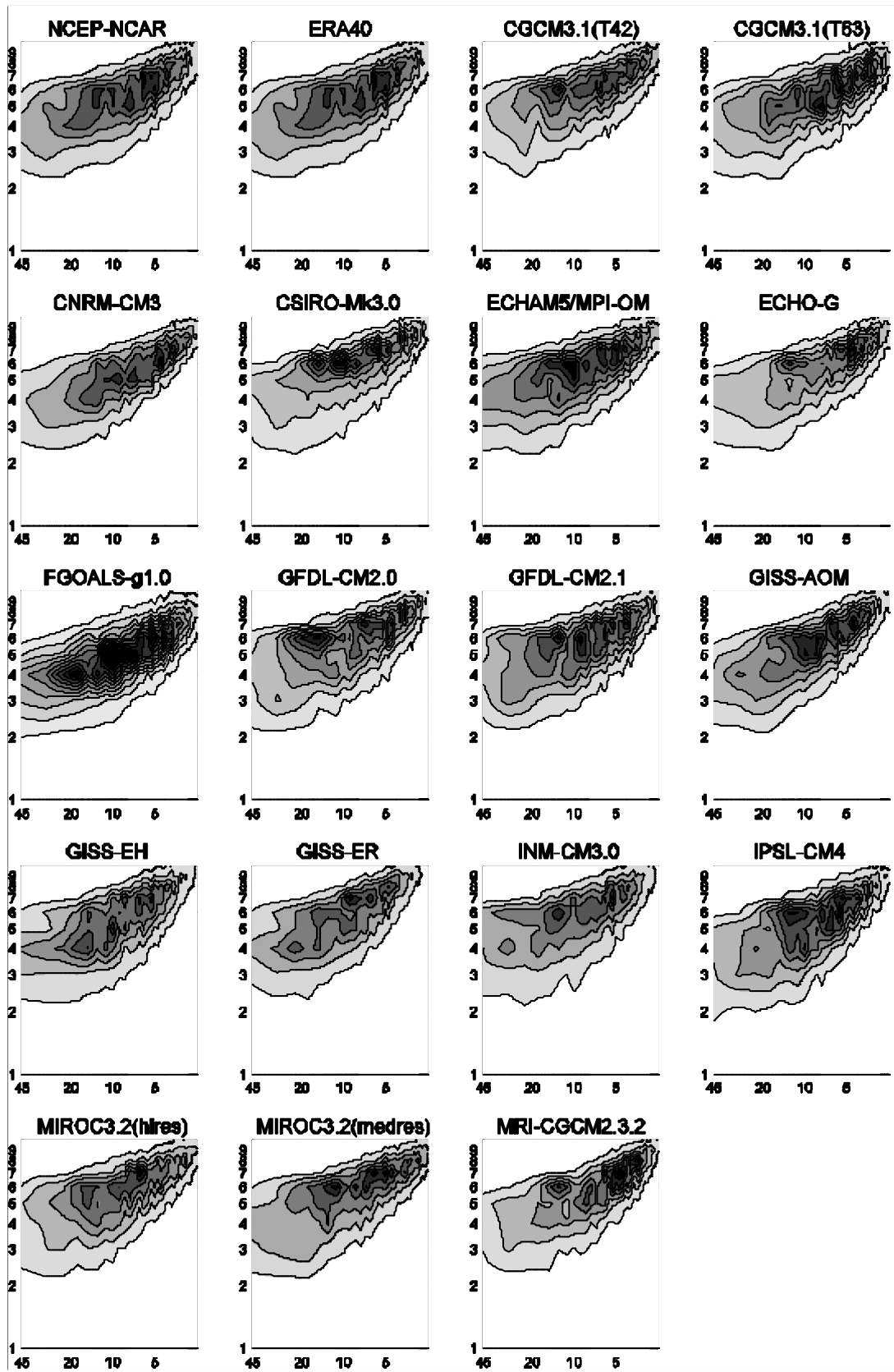

Figure 2: As in figure 3 but for the eastward propagating component.

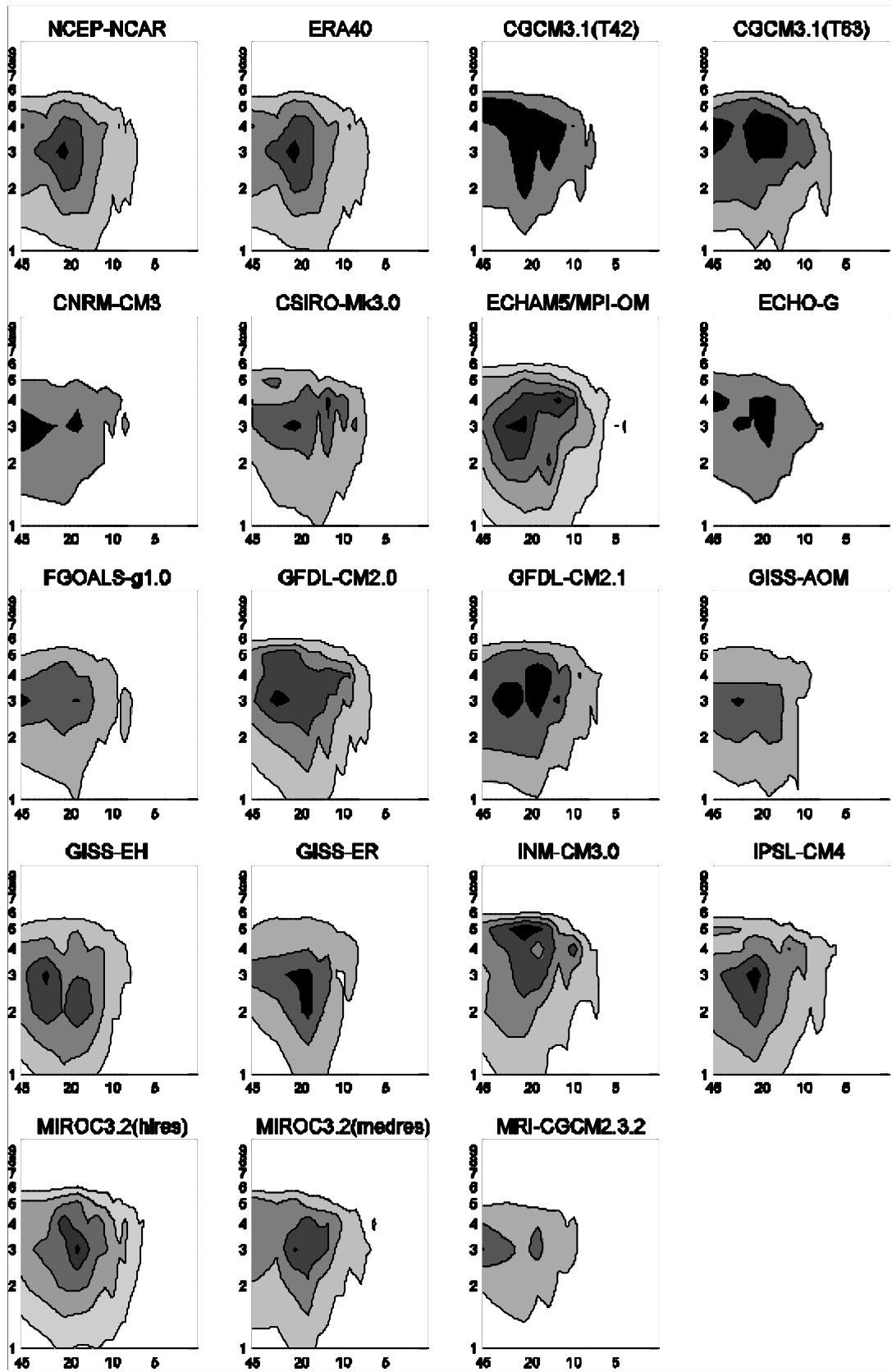

Figure 5: As in figure 3 but for the westward propagating component.

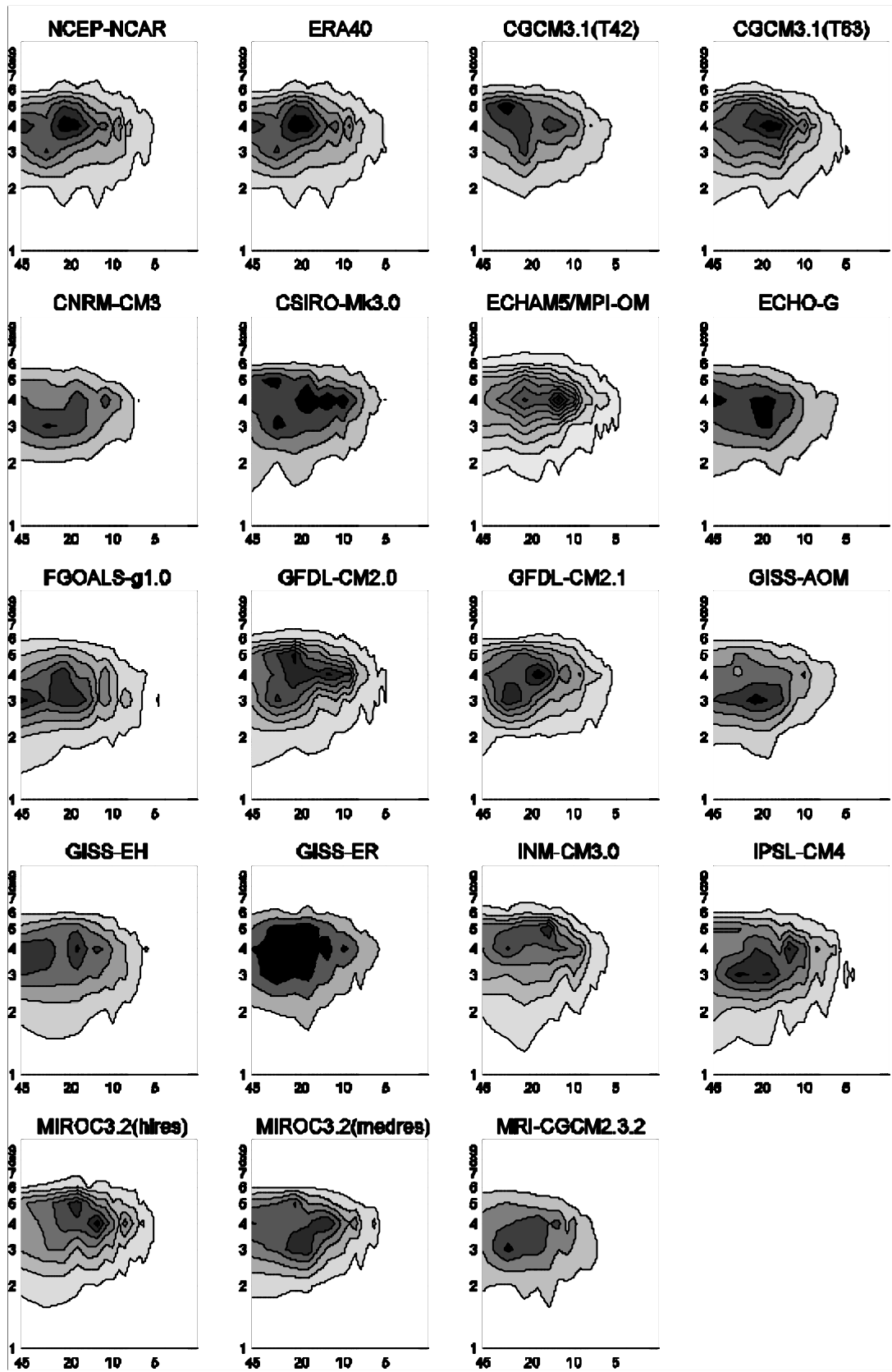

Figure 6: As in figure 3 but for the standing component.

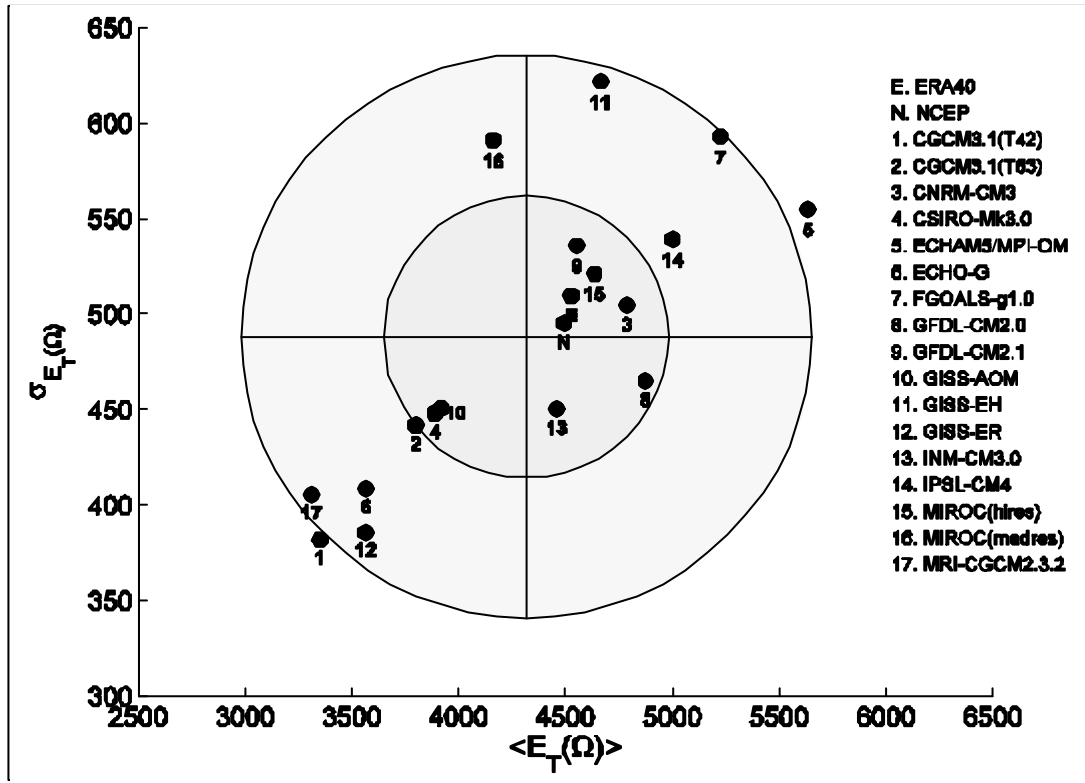

Figure 7: Mean annual full spectrum integral $\overline{E}_T(\Omega)$ and its variance for the 2 reanalyses and the 17 GCMs. The letters indicate the data computed from the NCEP-NCAR reanalysis (N), from the ERA40 reanalysis (E). The shaded areas represent the dispersion of data: the center of the ellipses is the ensemble average; the semi-axes of the inner ellipse are equal to the variance of data in the corresponding direction; the semi-axes of the outer ellipse correspond to twice the variance.

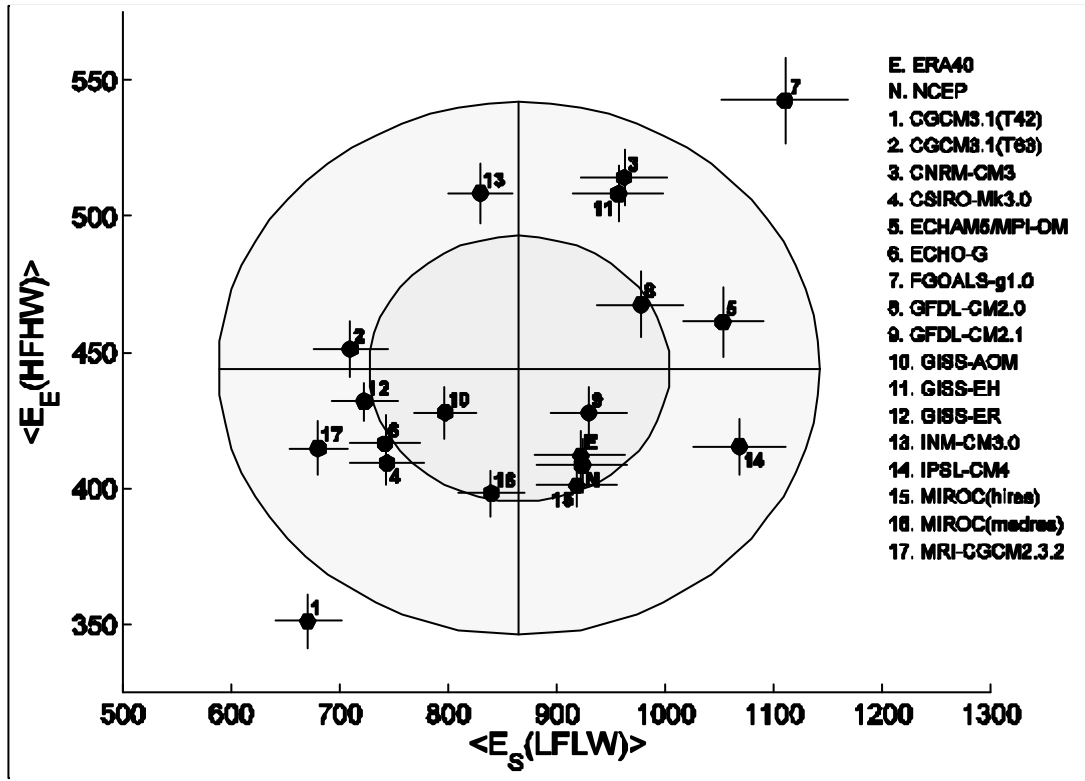

Figure 8: As in figure 7 but for the low-frequency low-wavenumber subdomain $\overline{E}_S(\Omega_{LFLW})$ (abscissas) of the standing waves versus the high-frequency high-wavenumber subdomain $\overline{E}_E(\Omega_{HFHW})$ (ordinates) of the eastward propagating waves. The crosses centered on